\documentclass[twocolumn,twoside,slac]{revtex4}
\usepackage{graphicx}
\usepackage{fancyhdr}
\begin{document}

\title{GRAPPA: Grid Access Portal for Physics Applications}

\author{Daniel Engh}
\affiliation{University of Chicago, Chicago, IL}
\author{Shava Smallen}
\affiliation{SDSC, San Diego, CA,}
\author{Jerry Gieraltowski}
\affiliation{Argonne National Lab, Chicago, IL}
\author{Robert Gardner}
\affiliation{University of Chicago, Chicago, IL}
\author{Liang Fang}
\affiliation{Indiana University, Bloomington, IN}
\author{Dennis Gannon}
\affiliation{Indiana University, Bloomington, IN}
\author{Randy Bramley}
\affiliation{Indiana University, Bloomington, IN}

\begin{abstract}
Grappa is a Grid portal effort designed to provide physicists
convenient access to Grid tools and services. The ATLAS analysis and
control framework, Athena, was used as the target application.  Grappa
provides basic Grid functionality such as resource configuration,
credential testing, job submission, job monitoring, results
monitoring, and preliminary integration with the ATLAS replica catalog
system, MAGDA.  Grappa uses Jython to combine the ease of scripting
with the power of java-based toolkits.  This provides a powerful
framework for accessing diverse Grid resources with uniform
interfaces.  The initial prototype system was based on the XCAT
Science Portal developed at the Indiana University Extreme Computing
Lab and was demonstrated by running Monte Carlo production on the
U.S. ATLAS test-bed.  The portal also communicated with a European
resource broker on WorldGrid as part of the joint iVDGL-DataTAG
interoperability project for the IST2002 and SC2002 demonstrations.
The current prototype replaces the XCAT Science Portal with an xbooks
jetspeed portlet for managing user scripts.
\end{abstract}

\maketitle

\thispagestyle{fancy}


\section{Introduction}

As the computational demands of High-Energy Physics (HEP) rises,
deploying physics analyses across the computational
``Grid''\cite{gridbook,globusweb} promises to meet much of this
demand.  However, Grid-enabling physics applications adds the
complexity of using a wide array of computing systems that differ in
their specific configurations and usage policies.  We have developed
Grappa:{\em Grid portal access for physics applications}\cite{grappa},
a Grid portal that provides a user-friendly approach for performing
tasks on the Grid such as: submitting and monitoring jobs, monitoring
system performance, and browsing results.

CPUs, Data Storage, software libraries are distributed across the
Grid.  A Grid portal, accessible from any browser, consolidates access
to these resources with uniform interfaces for the end-user who can
then think in higher-level terms such as total computational need or
data sets rather than individual CPUs or files, for example.

Physicists are typically familiar with HTML for presenting static
information, but they are typically less familiar with methods to
develop portals with more advanced functionality.  Grappa provides to
the physicist a modern, user-friendly framework for constructing web
portals with HTML forms and commonly understood scripting languages.

The current Grappa prototype uses the scripting language
Jython\cite{jython} with the Java Commodity Grid (Java
CoG)\cite{cog,cogweb} providing Grid tools.  With pre-configured Jython
wrapper functions to commonly used Java tools, a physicist-user can
readily develop customized portals with a minimum of Java programming
experience.

\section{The Portal Framework}

Grappa has undergone several revisions since the initial prototypes
began in the fall of 2001.  The initial Grappa portal was based on the
XCATSP\cite{xcatsp} with demonstrations of several successive
notebooks (and xbooks) for ATLAS\cite{atlas} job submission.  Other
groups have developed portals with features similar to Grappa, but
portals based on differing underlying frameworks have incompatible
components, hindering the ability to interchange these portal
components.  Jetspeed\cite{jetspeed} has been widely adopted as a
standard framework for portal components (portlets), allowing the
development of portlets with generic functionality that can be shared
among many different specialties (physics, chemistry, etc.).  Several
portlets have been developed by the Indiana University Extreme
Computing Lab\cite{extreme} and incorporated into Grappa.

The current Grappa prototype uses four component layers: Tomcat,
Jetspeed, Xbooks, and the user's xbook which consists of
scripts and HTML forms.  Our Grid portal architecture uses Java
"Portlet" technology to support secure interaction with Globus and
Grid web Services.  Each portlet defines a window on a Grid service or
instances of a Grid application using the Java CoG kit.  Users can
customize their own portal layouts by choosing and organizing
different portlets.  Grappa is a deployment of portlet combinations
including the xbooks portlet, tailored to some of the specific needs
for HEP data analyses.

\begin{figure*}[t]
\centering
\includegraphics[width=120mm]{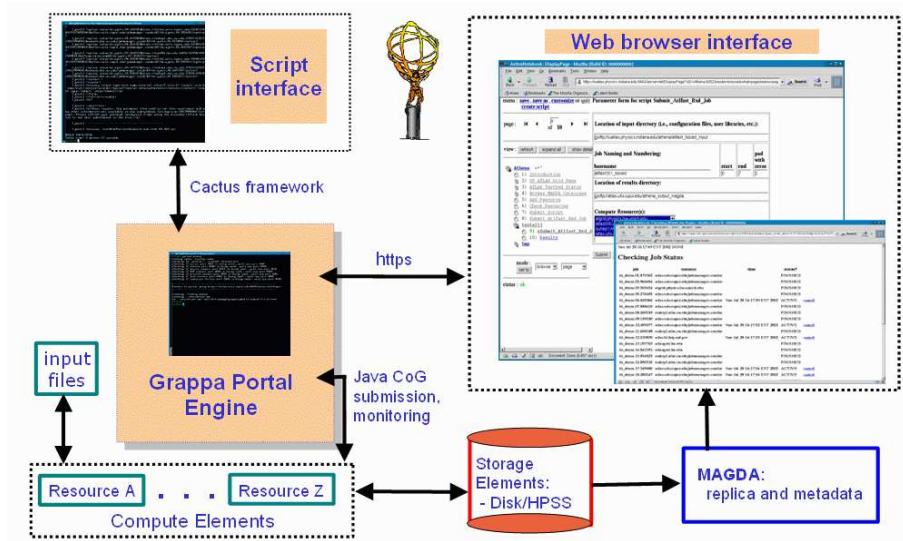}
\caption{Grappa job submission schematic.  A web or command-line
interface communicates with the portal which interacts with grid
resources.} \label{grappaSchematic}
\end{figure*}

\subsection{Xbooks}

Xbooks\cite{xbooks} provides the engine for interpreting the xbook
scripts, which in our case use Jython.  To use xbooks, a portal
(e.g. Grappa) is configured with an xbooks portlet.  This portlet
contacts an xbooks directory service and the user can choose a
particular xbook to run.  The xbook portlet finds an xbook server that
hosts the xbook, and this server sends the portlet a HTML form.  The
user-filled form values are sent to the xbooks manager which archives
them and uses them to configure and launch the application.  The
archival of the job input and submission values allows the user to
review, monitor, and even re-submit the jobs.

\subsection{Scripting}

Xbooks supports multiple scripting languages such as Python, Jython,
and Perl.  Grappa uses Jython to leverage the availability of Java
toolkits.  Jython wrapper functions to Java CoG tools provide easy
scripting interfaces for gsiftp file transfer, GRAM job submission and
GRAM job monitoring.  The xbook-developer creates HTML form
interfaces for entering information and scripts for processing this
information.  These scripts are split into 2 groups: scripts that run
on the portal host, and scripts that run on the compute host.

Before jobs are submitted onto a Grid-enabled cluster, Grappa makes
sure an updated cache of Grappa tools is available on that cluster.
These tools are packaged in a jar file that includes the Java CoG kit
and a core set of Grappa Jython scripts.  Grappa then submits jobs to
the Globus job-manager available for that system.  When these jobs
run, they then have access to the Grappa-installed Java tools which
include the Java CoG Globus client tools for gsiftp file transfer,
etc.

Jobs can be configured to search for locally installed libraries or to
temporarily install these libraries if needed, in our case the ATLAS
Athena libraries.  Application files, input data, and executables are
transferred from the Grid location specified on the web form.  The job
executes, transfers results to the results location, and finally
cleans up after itself.

\section{Prototypes}

The basic Grappa prototype is illustrated in \mbox{Figure
\ref{grappaSchematic}}.  The user interacts with the portal via web
browser or command-line interfaces.  The portal scripts use Java CoG
to submit to grid resources.  The portal can monitor GRAM information
for each job and the web interface can be used to monitor portal job
'side-effects' such as its MAGDA file registration.

We have developed and tested several Grappa prototypes.  The initial
prototypes, based on the XCATSP demonstrated resource management, job
submission, job monitoring, and browsing of results.  These features
were demonstrated in ATLAS Monte Carlo data challenges on U.S. ATLAS
test-beds.  The current prototype replaces XCATSP with Jetspeed
portlets, but uses essentially the same Jython scripts and HTML forms
from previous versions.  With the exception of the resource management
forms, the Grappa scripts have been successfully ported to xbooks and
the new Jetspeed-based portal demonstrates significantly improved
flexibility in portal configuration.

\subsection{Resource Management}

Information on the cluster, Grid, user, and application installations
for a compute resource is needed to submit jobs on that resource.  The
current model for managing this data in Grappa is to enter this
information on a portal web form, and this information is then stored
in a portal database.  Compute-cluster information stored in the
database includes the operating system and Java versions, the number
of processors on that cluster, and firewall information.  Grid
information for a resource includes the type of job manager, the
Globus job-manager contact string (as specified in the Globus GRAM
specification), and the gsiftp or gridftp server contact
strings. Application-specific information, such as the location of
application libraries, may also be entered in the resource database.
Application-specific information, however, is optional for the
resource database since the portal is designed with the flexibility to
be able to run applications on generic systems that do not have any
application-specific software pre-installed.  An improved model for
managing this data would be to dynamically access resource information
via MDS\cite{mds}.

The user submits a second resource management web form to test the
availability of selected resources.  Grappa uses Globus authentication
to restrict portal access and provide access by proxy to grid
resources.  The initial model used credentials obtained from the
.globus directory of the user that instantiated the portal.  A future
model to use a Jetspeed portlet to manage proxies obtained from a
MyProxy Server has also been demonstrated.

\subsection{Job Submission}

The user specifies additional information on a job submission web
form.  This information includes locations for input and output data,
user-specific applications and libraries, and (for example)
collaboration-certified library packages.  The user enters these
locations with the Uniform Resource Identifier (URI) format
(protocol/host/path) which generalizes paths to the Grid level, so
these locations can be scattered across the Grid while the user thinks
in terms of a single form.  The portal scripts use this information to
dynamically construct a working environment on generic Grid-enabled
compute sites.

Additional application-specific information, such as the number of
events and the physics model to simulate are entered and passed as
arguments from the portal to the application scripts (e.g. jobOptions
files).

Rather than submitting jobs to specific sites, the user operates on a
higher level and defines an active {\em set} of compute resources
selected from the compute resources--previously entered into the
compute resource database and checked for user accessibility.  Grappa
submits jobs to this set of resources and, transparent to the user,
the number of jobs submitted to each site is proportioned to the
relative computing power of that site.

\begin{figure*}[t]
\centering
\includegraphics[width=150mm]{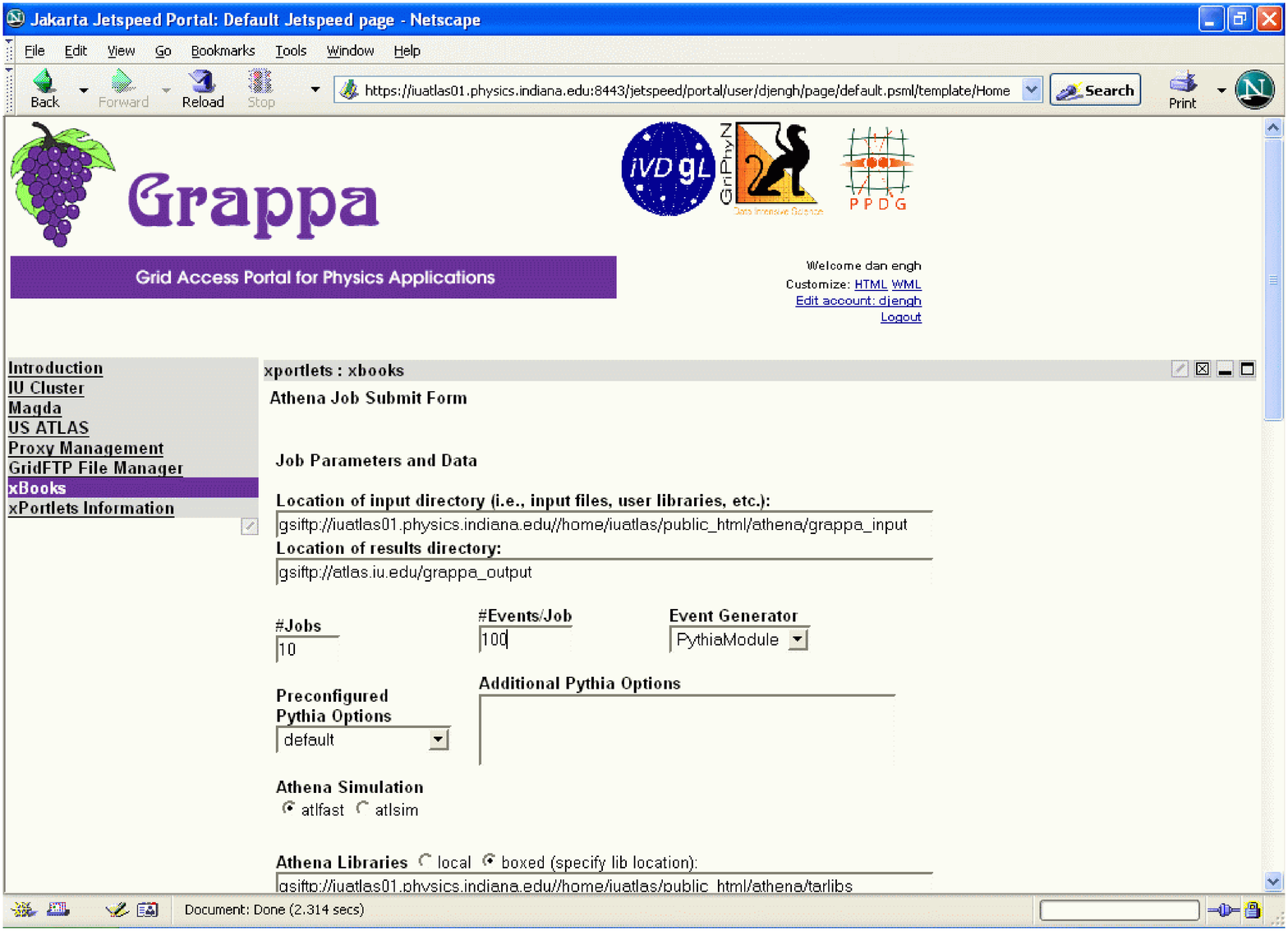}
\caption{A screen-shot showing a portion of the Grappa submission
form.  The user enters grid locations for input and output files, job
parameters, etc.} \label{GrappaSubmit}
\end{figure*}

A portion of the job submission form is visible in Figure
\ref{GrappaSubmit}

\subsection{Job Monitoring}

\begin{figure*}[t]
\centering
\includegraphics[width=150mm]{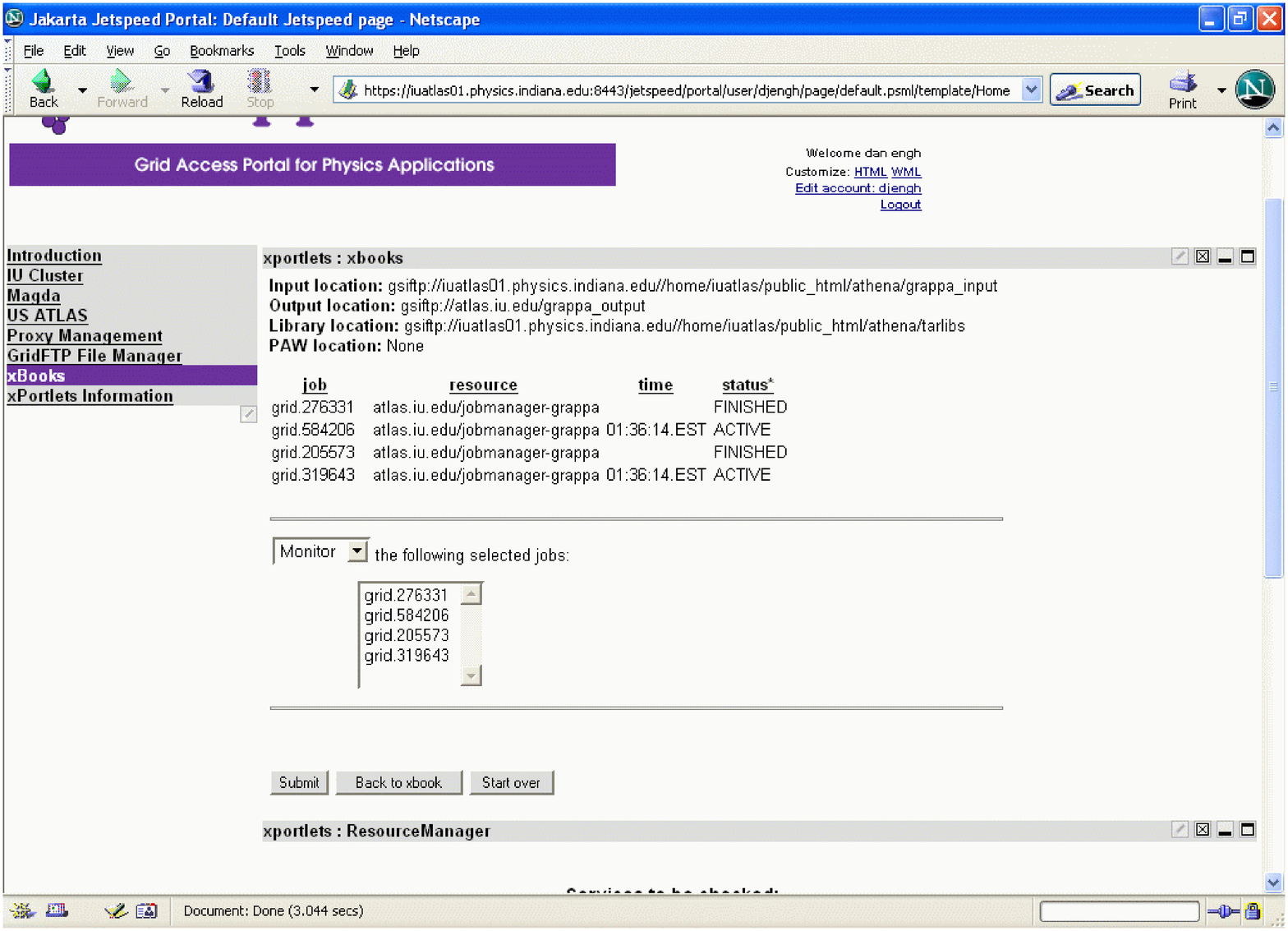}
\caption{Grappa screen-shot showing job submission results.  Four jobs
have been submitted and their GRAM status reported.  The user can
choose to further monitor or cancel jobs.}
\label{GrappaMonitor}
\end{figure*}

The user monitors Grappa jobs in two ways.  Submitted jobs appear on a
web form.  From this, the user selects lists of jobs to monitor.
Grappa then queries the GRAM reporters on the compute sites to obtain
the status for each job.  A screen-shot of the Grappa job monitoring
form is shown in Figure \ref{GrappaMonitor}.  Links of system monitors
such as Ganglia were added to the portal to provide a second method
to monitor Grappa jobs and cluster performance.  Additionally links
to the ATLAS replica catalog system, MAGDA\cite{magda}, provide the
user the ability to browse file locations from within the portal.

\subsection{Browsing Results}

Visual results of ATLAS Monte Carlo ntuples were generated with PAW.
The user enters locations for a PAW package and a user PAW macro
(kumac) into a web form.  Grappa then installs and runs PAW on the
ntuples as they are created.  Visual monitors of individual jobs are
available upon completion of each job.  In addition a summary plot
dynamically updates with the results of the entire multi-file dataset
as it is being generated.  The Grappa-produced graphics files and log
files can be browsed using a GridFTP File Manager Jetspeed portlet.

\subsection{Data Challenges}

The Grappa framework is packaged and installed on submit hosts with
PACMAN\cite{pacman}.  As the target application we used the
ATLAS\cite{atlas} analysis and control framework, Athena, configured
to run a fast Monte Carlo simulation, Atlfast.  Athena libraries and
PAW are available as tar files that Grappa dynamically installs on any
compute site as needed.

The portal is designed to submit to Globus-supported job-mangers.  The
U.S. test-bed used for Grappa data challenges provided access to about
100 CPUs across 15 different sites using condor, lsf, pbs, and fork
Globus job-managers.  The U.S. ATLAS grid test-bed included
contributions from the Universities of: Texas (Arlington), Oklahoma,
Chicago, Indiana, Michigan, Wisconsin, Florida, and Boston, plus
Fermi, Argonne, Berkeley, and Brookhaven National Labs.  To
demonstrate interoperability with a European resource broker, the
portal communicated with the resource broker (INFN) using the Globus
fork job-manager, but the application was packaged to provide a Job
Description Language (JDL) script to the broker which then resubmits
the job to its own resources.

Grappa performance is limited by the wait time between each job
submission.  The delays in contacting remote resources using
Grappa/Java Cog were similar to the delays seen using Globus via a
Unix Shell or Condor-G\cite{condorg}.  The long wait that occurs when
large numbers of jobs (greater than 25) resulted in a ``busy'' web
form for longer than 10 minutes, making these large scale submissions
somewhat impractical from an interactive web form.  This problem was
solved by creating a non-interactive command line interface to the
portal using the Cactus\cite{cactus} toolkit.  For large-scale
long-duration data production, cron is used to submit the desired
frequency of jobs.  This saves the web interface from long wait times
so the status and progress of web-submitted and command-line-submitted
jobs can then be viewed from the web interface whenever the user
wishes.

\section{Summary}

Grid Portals promise to simplify for the end-user access to diverse
Grid resources.  Grappa provides a powerful set of tools
allowing development of platform independent user-interfaces
accessible via standard Internet protocols.  We have demonstrated with
ATLAS Monte Carlo data challenges that Grappa can be a useful
interface for controlling large scale data production on the Grid.\\

\section{Acknowledgments}
This work was supported in part by the National Science Foundation.

\bibliographystyle{unsrt}
\bibliography{grappa}

\begin{thebibliography}{10}

\bibitem{gridbook}
Ian Foster and Carl Kesselman, editors.
\newblock {\em {The Grid: Blueprint for a New Computing Infrastructure}}.
\newblock {Morgan Kaufmann Publishers, Inc.}, San Francisco, USA, 1999.

\bibitem{globusweb}
{Globus} webpage at \url{http://www.globus.org}.

\bibitem{grappa}
{Grappa} webpage at \url{http://grid.uchicago.edu/grappa/}.

\bibitem{jython}
{Jython} webpage at \url{http://www.jython.org}.

\bibitem{cog}
Gregor von Laszewski, Ian Foster, and Jarek Gawor.
\newblock {CoG Kits: A Bridge Between Commodity Distributed Computing and
  High-Performance Grids}.
\newblock In {\em ACM 2000 Java Grande Conference}, June 2000.

\bibitem{cogweb}
{JavaCog} webpage at \url{http://www.globus.org/cog}.

\bibitem{xcatsp}
Sriram Krishnan, Randall Bramley, Dennis Gannon, Madhusudhan Govindaraju, Rahul
  Indurkar, Aleksander Slominski, Benjamin Temko, Jay Alameda, Richard Alkire,
  Timothy Drews, and Eric Webb.
\newblock { The XCAT Science Portal }.
\newblock In {\em { Proceedings of Supercomputing 2001}}, 2001.

\bibitem{atlas}
{ATLAS} webpage at \url{http://atlasexperiment.org}.

\bibitem{jetspeed}
{Jakarta Jetspeed} webpage at \url{http://jakarta.apache.org/jetspeed }.

\bibitem{extreme}
{Extreme! Computing Lab} webpage at \url{http://www.extreme.indiana.edu}.

\bibitem{xbooks}
{Xbooks} webpage at \url{http://www.extreme.indiana.edu/xbooks}.

\bibitem{mds}
K.~Czajkowski, S.~Fitzgerald, I.~Foster, and C.~Kesselman.
\newblock {Grid Information Services for Distributed Resource Sharing}.
\newblock In {\em Proceedings of the Tenth IEEE International Symposium on
  High-Performance Distributed Computing (HPDC-10)}. IEEE Press, August 2001.

\bibitem{magda}
{MAGDA} webpage at \url{http://www.atlasgrid.bnl.gov/magda/info}.

\bibitem{pacman}
{PACMAN} webpage at \url{http://physics.bu.edu/~youssef/pacman/}.

\bibitem{condorg}
{Condor-G} webpage at \url{http://www.cs.wisc.edu/condor/condorg/}.

\bibitem{cactus}
{Cactus} webpage at \url{http://jakarta.apache.org/cactus}.

\end{thebibliography}

\end{document}